\documentclass{article}

\usepackage[preprint, nonatbib]{neurips_2022}

\usepackage[utf8]{inputenc} % allow utf-8 input
\usepackage[T1]{fontenc}    % use 8-bit T1 fonts
\usepackage{hyperref}       % hyperlinks
\usepackage{url}            % simple URL typesetting
\usepackage{booktabs}       % professional-quality tables
\usepackage{amsfonts}       % blackboard math symbols
\usepackage{nicefrac}       % compact symbols for 1/2, etc.
\usepackage{microtype}      % microtypography
\usepackage{lipsum}
\usepackage{graphicx}
\usepackage{subfigure}
\graphicspath{ {./images/} }
\usepackage{xspace} 
\usepackage{amsmath}
\usepackage{amssymb}
\usepackage{csquotes}
\begin{document}

\title{Gabor-based learnable sparse representation for self-supervised denoising}
\author{
 Sixiu Liu, Shijun Cheng, Tariq Alkhalifah \\
  King Abdullah University of Science and Technology, \\
  \texttt{sixiu.liu@kaust.edu.sa, sjcheng.academic@gmail.com,} \\
  \texttt{tariq.alkhalifah@kaust.edu.sa} \\
}
\date{} % clear date
\maketitle

\begin{abstract}
Traditional supervised denoising networks learn network weights through “black box” (pixel-oriented) training, which requires clean training labels. The uninterpretability nature of such denoising networks in addition to the requirement for clean data as labels limits their applicability in real case scenarios. Deep unfolding methods unroll an optimization process into Deep Neural Networks (DNNs), improving the interpretability of networks. Also, modifiable filters in DNNs allow us to embed the physics information of the desired signals to be extracted, in order to remove noise in a self-supervised manner. Thus, we propose a Gabor-based learnable sparse representation network to suppress different noise types in a self-supervised fashion through constraints/bounds applied to the parameters of the Gabor filters of the network during the training stage. The effectiveness of the proposed method was demonstrated on two noise type examples, pseudo-random noise and ground roll, on synthetic and real data.
\end{abstract}

% \par
% \textbf{Key words:} Self-supervised, denoise, CNNs, seismic processing
% keywords can be removed

\section{Introduction}
Seismic data are often composed of useful signals, but also noise that can be classified into incoherent and coherent. High levels of noise compared to the signal can negatively impact subsequent inversion and imaging applications \cite{yilmaz2001seismic}. Incoherent noise, or random noise, potentially caused by instrumentation issues or the environment, refers to the seismic energy that is uncorrelated along the time or space axis, while coherent noise represents the energy that is correlated along the time or/and space axis, such as trace-wise noise, bandpassed noise and ground roll in reflection seismology. Ground roll, which represents surface waves in land data, does not contribute to our objectives in imaging reflection energy and is of a coherent nature, so we regard it as noise \cite{mcmechan1991depth,henley2003coherent,halliday2015scattered}. The general dispersive property in which different frequency components of the Rayleigh waves travel at different velocities results in the ground roll having a \enquote{fan} like shape in seismic data, with distinctive slopes representing the low velocity nature of the Rayleigh waves  \cite{beresford1988dispersive,deighan1997ground}. Thus, ground roll is often characterized by its high amplitude, low velocity, low frequency, and is often spatially aliased, masking seismic signals and reducing the Signal to
Noise Ratio (SNR) in our seismic data \cite{claerbout1985ground}. Therefore, ground roll attenuation is a crucial step in the early stage of seismic data processing to enhance the data quality and obtain accurate inversion results \cite{claerbout1985ground,saatcilar1988method,henley2003coherent}.

Various methods have been proposed for removing seismic noise. For example, prediction-based methods suppress noise by assuming signals are predictable but noise is not, such as the Polynomial Fitting (PF) \cite{lu2006local,liu2011seismic} and f-x deconvolution \cite{canales1984random}. Decomposition-based approaches reconstruct signals from the main components of noisy data, which is assumed to be decomposed into different components, such as Singular Value Decomposition (SVD) \cite{kendall2005svd,cary2009ground,porsani2009ground} and Empirical Mode Decomposition (EMD) \cite{bekara2009random,gomez2016simple}. 

Conversely, sparse transform, which assumes signals can be sparsely represented and approximated by linearly combining a few basis functions, has emerged as a more popular technique for noise removal in seismic data. This method involves transforming the noisy data into a sparse representation in a specific domain, where a threshold is used to filter out the noise. The filtered representations are then transformed back to the original domain. The sparse representation can be obtained using either analytic or learning-based methods \cite{tang2012seismic}. Analytic methods typically employ a fixed set of basis functions, for F-K filtering \cite{foti2002spatial}, Fourier transform \cite{yilmaz2001seismic}, radon transform \cite{liu20043,trad2001hybrid,hu2016ground}, shearlet transform \cite{hosseini2015adaptive}, wavelet transform \cite{deighan1997ground,iranimehr2021seismic,lin2022ground}, curvelet transform \cite{yarham2006curvelet,naghizadeh2018ground,wang2020ground}, and S transform \cite{tan2013ground,tao2019second}, to fit the signals. Although sparse transforms based on fixed functions are computationally efficient and can be applied in real-time scenarios, they rely on the assumption that the seismic noise can be represented by a fixed set of functions that are different than those used to represent the signal, which may not always be the case. In other words, fixed transforms are often not able to fully separate signals from noise, leading to some loss of detail or resolution in the denoised output \cite{birnie2021potential}.

Alternatively, learning-based approaches can learn, possibly optimal for the task and data, basis functions using a training process. A representative approach is Dictionary Learning (DL), which enables us to learn adaptive basis functions, known as dictionaries, from training data. Each dictionary consists of a set of \enquote{atoms}. Using a few atoms in a learned dictionary, we can sparsely represent signals. DL has been widely explored in seismic signal processing, such as denoising \cite{harsuko2022storseismic,chen2016double,zhou2020statistics}, interpolation \cite{lan2021robust,cheng2023meta,tang2023simultaneous}, imaging \cite{cheng2023elastic}, and inversion \cite{shao2021simultaneous, du2022deep,alali2023integrating}. One popular example of DL is K-Singular Value Decomposition (K-SVD) \cite{aharonm2006thek}, which iteratively updates a dictionary using SVD and uses the dictionary to obtain sparse representations. K-SVD has been widely used in seismic denoising \cite{cheng2018seismic,chen2020fast,chen2022retrieving} and interpolation \cite{hou2018simultaneous}. However, K-SVD updates one single element of the dictionary based on overlapping small patches, failing to capture global features in the data \cite{turquais2017method}. Also, the SVD  operations in K-SVD are computationally expensive, which limits its use on large datasets. Other DL-based methods include Data-Driven Tight Frame (DDTF) \cite{yu2015interpolation,nazari2017data,wang2019adaptive}, double-sparsity dictionary \cite{chen2016double,zhu2017joint}, multiscale DL scheme \cite{zhu2015seismic}, graph-regularized DL method \cite{liu2018sparse}, dip-oriented DL method \cite{zu2019dictionary}, coherence-based DL method \cite{turquais2017method,wang2019adaptive}, and transform learning \cite{wang2019denoising}.

While DL suppresses noise by representing a given image through linear combinations of the learned atoms in a dictionary, Deep Neural Networks (DNNs) aim to remove noise by extracting data features through learned weights or filters. DNNs are becoming popular in seismic noise suppression tasks, such as random noise removal \cite{mandelli2019interpolation,liu2018random,li2021deep} and coherent noise removal \cite{li2019ground,oliveira2020ground,pham2022physics}. However, most of the Deep Neural Network (DNN) models regard a learned network as a \enquote{black box} without feeding prior knowledge about signals or noise to the network, meaning the learned functions by the network are not interpretable and not completely understood by humans. Different from traditional DNNs that are not interpretable, deep unfolding methods \cite{gregor2010learning} unfold iterative optimization algorithms into DNNs and are interpretable due to the combination of DNN models and optimization algorithms. Deep unfolding methods are utilized in computer vision tasks \cite{wisdom2017building,zhang2020deep,monga2021algorithm}, such as image denoising \cite{barbu2009training,sun2011learning,chen2016trainable} and image Compressive Sensing (CS) \cite{sun2016deep,zhang2018ista}. In \cite{zhang2018ista}, an Iterative Shrinkage Threshold Algorithm-based Network (ISTA-Net) embeds an Iterative Shrinkage
Threshold Algorithm (ISTA) \cite{beck2009fast} into DNNs for CS reconstruction of natural images. Similarly, a Convolutional Dictionary Learning Network (CDLNet) \cite{janjuvsevic2021cdlnet} provides a robust and interpretable network for natural image denoising by incorporating ISTA into DNNs. 
Due to the \enquote{Gabor-like} dictionaries in CDLNet,  a Gabor-based Dictionary Learning Network (GDLNet) \cite{janjuvsevicc2022gabor} replaces the learnable convolutional filters of CDLNet by parameterized learnable Gabor functions defined by much fewer learnable parameters, and yet obtains comparable image denoising performance. For seismic data, \cite{sui2023deep} used an interpretable Deep Unfolding Dictionary Learning (DUDL) method that unrolls an iterative algorithm of DL into DNNs for seismic denoising. However, the method is supervised, requiring clean labels that are often not available for seismic field records.

Recently, inspired by Noise2Void (N2V) \cite{krull2019noise2void} and  Structured Noise2Void (STRUCTN2V) \cite{broaddus2020removing} in natural imaging denoising tasks, self-supervised methods are gaining momentum in seismic denoising tasks as they train the network only on noisy data, without requiring clean versions \cite{liu2022self,liu2022aiding,birnie2023explainable}. Self-supervised schemes can be classified into two types: input masking methods and network masking methods. Input masking methods pre-process noisy data as inputs for network training, such as those for seismic random \cite{birnie2021self,birnie2021potential,sun2022seismic,cao2022self} and coherent noise removal \cite{liu2022coherent,liu2022self,abedi2022multidirectional}. In contrast,  network masking methods modify the receptive field of the network to implement deblending in a self-supervised fashion \cite{luiken2023integrating,wang2022self}. Inspired by network masking methods where the receptive field of a network is restricted according to the noise property, the design of the filters can also be investigated in terms of the property of targeted signals.

However, the suppressed noise by the aforementioned self-supervised procedures is usually of a single type and involves simple noise scenarios, such as random, trace-wise noise, or bandpassed noise, instead of a batch of different noise types. In this study, based on the interpretation property and the designable Gabor filters of GDLNet, we adapt the original GDLNet, to seismic denoising by incorporating the domain knowledge of the desired signals into the design of the network's Gabor filters. The proposed method can remove different noise types, such as pseudo-random noise and ground roll, in a self-supervised fashion with far fewer epochs of training. Addressing different noise types only require a subtle parameter setting change of the initial  Gabor filters of the network, therefore the proposed procedure is potentially robust in handling many noise types.

This study is organized as follows. We first review the theory of DL, followed by the original GDLNet architecture, and the self-supervised denoising theory with GDLNet architecture. Finally, the proposed method is applied to suppress two noise types, pseudo-random noise and ground roll, on both synthetic and field data.

\section{Theory}
\subsection{Dictionary Learning (DL)}
DL is divided into two parts: updating the dictionary and updating the sparse representation. Assuming we have noisy data $\mathbf{y}$ and clean data $\mathbf{x}$, DL aims to reconstruct the clean data $\mathbf{x}$ from noisy data $\mathbf{y}$ by solving the following optimization problem \cite{sui2023deep}
\begin{equation}
\min _{\mathbf{D}, \mathbf{z}} \frac{1}{2}\|\mathbf{y}-\mathbf{D z}\|_2^2+\lambda_D R_1(\mathbf{D})+\lambda_z R_2(\mathbf{z}),
\end{equation}
where $\mathbf{y}$, $\mathbf{D}$, and $\mathbf{z}$ are the noisy data, the dictionary, and the sparse coefficient vector, respectively. $R_1$ and $R_2$ are the regularizers for $\mathbf{D}$ and $\mathbf{z}$, with regularization parameters 
$\lambda_D$ and $\lambda_z$, respectively. $\|\cdot\|_2$ represents the $\ell_2$ norm. 

The dictionary $\mathbf{D}$ and the sparse coefficient vector $\mathbf{z}$ are updated in an alternate manner in DL. Given the dictionary $\mathbf{D}$, the sparse coefficient vector $\mathbf{z}$ can be updated through the minimization subproblem
\begin{equation}
\min _{\mathbf{z}} \frac{1}{2}\|\mathbf{y}-\mathbf{D z}\|_2^2+\lambda_z R_2(\mathbf{z}),
\end{equation}
where $R_2$ can be a sparse regularization for a denoising task to constrain the sparsity of the data. ISTA \cite{beck2009fast} is one popular method to solve the optimization of the sparse coefficient vector $\mathbf{z}$ due to its simple implementation, which is given by
\begin{equation} 
\mathbf{z}^{(k+1)}:=\operatorname{ST}\left(\mathbf{z}^{(k)}-\eta \mathbf{D}^{\top}\left(\mathbf{D} \mathbf{z}^{(k)}-\mathbf{y}\right), \eta \lambda_z\right),
\end{equation}
where $
\mathrm{ST}(\mathbf{r}, \tau) =\operatorname{\emph{$sign$}}(\mathbf{r}) \emph{$max$} (0,|\mathbf{r}|-\tau)
$ is a Soft-Thresholding operator applied on $\mathbf{r}$, with a non-negative threshold $\tau$. $\eta$ is the step size of the regularization parameter $\lambda_z$. $k$ is the iteration index. ${\top}$, as a superscript, is the transpose operator.

Similarly, to update the dictionary $\mathbf{D}$ with fixed sparse coefficient vector $\mathbf{z}$, the following optimization can be solved
\begin{equation}
\min _{\mathbf{D}} \frac{1}{2}\|\mathbf{y}-\mathbf{D z}\|_2^2+\lambda_D R_1(\mathbf{D}),
\end{equation}
where $R_1$ can be the unit-norm constraint on dictionary $\mathbf{D}$. The optimization of the dictionary $\mathbf{D}$ can be solved by the  Optimal Directions Method (ODM) \cite{engan1999frame} or K-SVD \cite{aharon2006k}. 

The alternative updating of the DL resembles the behavior of DNNs, which inspires GDLNet to enhance the interpretability and robustness of natural image denoising tasks.
\subsection{The original GDLNet architecture}
We first introduce the original GDLNet architecture \cite{janjuvsevicc2022gabor} depicted in 
Figure \ref{fig1}a.  Assuming we have a noisy data sample $\mathbf{y}$ and a clean data sample $\mathbf{x}$, GDLNet aims to reconstruct the clean data $\mathbf{x}$ from noisy data $\mathbf{y}$ via casting the ISTA for updating the dictionary and the sparse representation into DNNs, as follows,
\begin{equation}
\hat{\mathbf{x}}=\mathbf{D} \mathbf{z}^{(K)}, \quad \mathbf{z}^{(0)}=\mathbf{0}, \quad k=0,1, \ldots, K-1,
\end{equation}
\begin{equation}
\mathbf{z}^{(k+1)}=\operatorname{ST}\left(\mathbf{z}^{(k)}-\mathbf{A}^{(k)^{\top}}\left(\mathbf{B}^{(k)} \mathbf{z}^{(k)}-\mathbf{y}\right), \tau^{(k)}\right),
\end{equation}
where $k$ and $K$ are the layer index and the total number of layers of the network, respectively. $\mathbf{z}^{(0)}$ is the initial input sparse coefficient vector, initialized with zero and updated through the network layers by training the network. $\hat{\mathbf{x}}$ is the optimal denoised image, which is a convolution between the synthesis filterbank of the last layer of the network, referred to as a convolutional dictionary $\mathbf{D}$, and the optimized sparse coefficient vector $\mathbf{z}^{(K)}$ at the last layer. $\mathbf{A}^{(k)}$ and $\mathbf{B}^{(k)}$ are analysis and synthesis filterbanks, respectively. For a certain layer $k$, the synthesis convolution operation is defined by $\mathbf{B z}=\sum_{m=1}^M b^m * z^m$, where ${b}^m $ and ${z}^m $ correspond to the ${m}$-th element of $\mathbf{B}$ and $\mathbf{z}$, respectively, with ${M}$ being the number of elements. The analysis convolution operation is defined by 
$\left(\mathbf{A}^{\top} \mathbf{x}\right)^m=\overline{a^m} * \mathbf{x}$, where ${a}^m$ is the ${m}$-th element of $\mathbf{A}$
and $\overline{a^m}$ is the reversal of ${a}^m$. ST is again, a Soft-Thresholding operator with a learnable non-negative threshold $\tau^{(k)}$, which is generally constant in DL. $\mathbf{z}^{(.)}$ is the sparse coefficient vector updated by each layer.\\
\subsection{Adapted GDLNet architecture via Parameterized Gabor Filterbanks}
\begin{figure*}[!t]
\centering
\includegraphics[width=5.3in,height=2.8in]{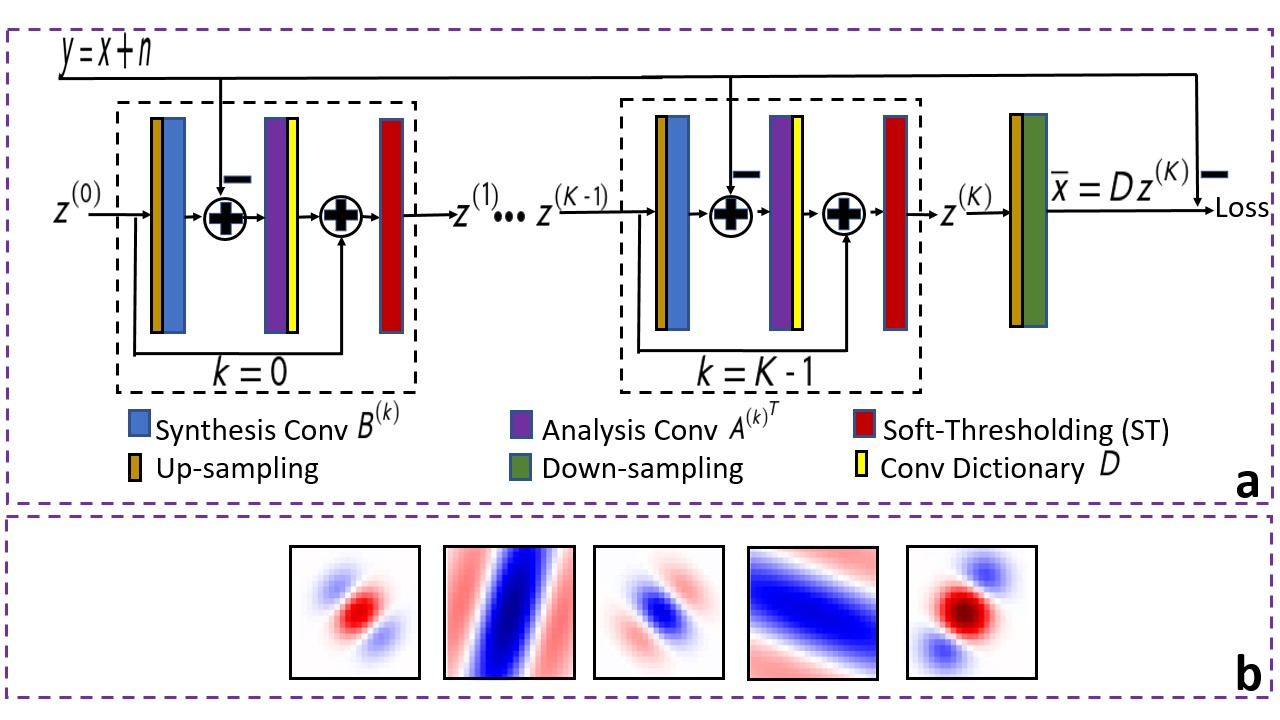}
% \captionsetup[figure]{font=Large}
\caption{a) The GDLNet architecture; b) Examples of the adapted Gabor filterbanks.}
\label{fig1}
\end{figure*}

The learned analysis filterbanks $\mathbf{A}^{(k)}$, synthesis filterbanks $\mathbf{B}^{(k)}$  and convolutional filterbanks $\mathbf{D}$ in GDLNet are \enquote{Gabor-like} filters and can be parameterized by learnable 2D real Gabor functions
\begin{equation} \label{eq:1}
g({x} ; \phi)=\alpha e^{-\|{a} \circ {x}\|_2^2} \cos \left({\omega}_0^{\top} {x}+\psi\right),
\end{equation}
where ${x}$ is the spatial position and $\circ$ represents elementwise product. $\phi=\left(\alpha, {a}, {\omega}_0, \psi\right)$ are the Gabor trainable parameters, in which $\alpha$ is the amplitude of the filters, ${a}$ is the (root) precision, ${\omega}_0$ is the spatial frequency, and $\psi$ is the phase offset of the cosine plane wave. The original Gabor filter in Equation \ref{eq:1} is not described by intuitive parameters, such as the direction of the cosine plane wave, which can be used to extract desired signals with specific directions. Therefore, we adapt a modefied Gabor function to represent localized Gaussian weighted plane waves defined by angle to replace that in Equation \ref{eq:1} in GDLNet, and it has the form:
\begin{equation} \label{eq:2}
g(x, y ; \alpha, \lambda, \theta, \psi, \sigma, \gamma)=\alpha \exp \left(-\frac{x^{\prime 2}+\gamma^2 y^{\prime 2}}{2 \sigma^2}\right) \cos \left(2 \pi \frac{x^{\prime}}{\lambda}+\psi\right), \\
\end{equation}
where $ x^{\prime}=x \cos \theta+y \sin \theta$ and $ y^{\prime}=-x \sin \theta+y \cos \theta$, with $\theta$ being the angle describing the orientation of the filter (the plane wave). $\alpha$ is the amplitude of the filter. $\lambda$ and $\psi$ represent the wavelength and phase offset of the cosine plane wave, respectively. $\sigma$ and $\gamma$ are the standard deviation and the spatial ratio of the transformed axis of the Gaussian function, respectively. Only six parameters are learned
to generate a Gabor filter of any size, which reduces the number of learnable parameters compared to a conventional convolutional filter of the same size. For example, to generate a convolutional filter of size $11\times11$, 121 parameters are required to be learned while only 6 parameters are learned by the Gabor filter. Some examples of the adapted Gabor filters are shown in Figure \ref{fig1}b. 
\subsection{Self-supervised training the modified GDLNet architecture }
Due to the six learnable and modifiable parameters of the Gabor filters in Equation \ref{eq:2}, constraints applied to the six parameters allow us to isolate certain characteristics of desired signals to be extracted from seismic data. The target and the input of the network are the same noisy data, that is, the network is trained in a self-supervised fashion. We expect the network will be able to learn targeted signals faster. 

To accomplish this, we employ learnable Gabor filters, which are given by Equation \ref{eq:2}, instead of conventional convolutional kernels in the network. Meanwhile, we leverage prior knowledge about the desired signals to initialize and impose constraints on the filter parameters, optimizing their effectiveness in capturing the underlying signal characteristics. To be specific, to generate suitable initial Gabor filters, the $\theta$ is randomly initialized from a range [$\theta_1$,$\theta_2$], where $\theta_1$ and $\theta_2$ are decided by the minimum and maximum slopes of the desired signals, respectively. Similarly, the wavelength $\lambda$ is also randomly initialized from the range [$\lambda_1$,$\lambda_2$], where $\lambda_1$ and $\lambda_2$ are decided by the minimum and maximum wavelength of the desired signals, with a unit of grid points, respectively. 

\begin{figure*}[!t]
\centering
\includegraphics[width=5.4in]{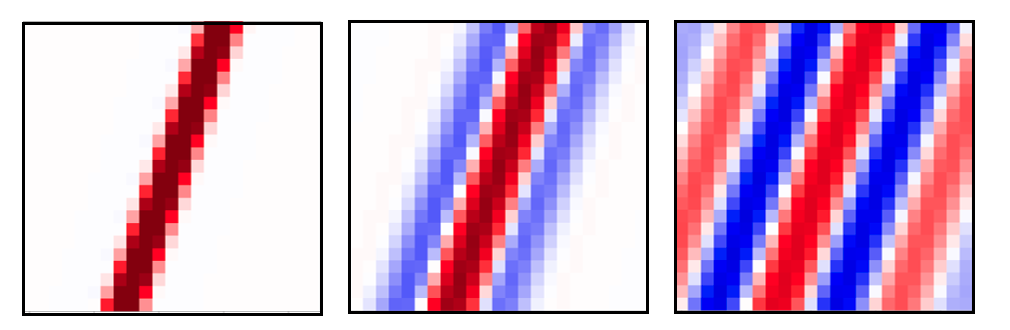}
% \captionsetup[figure]{font=Large}
\caption{Gabor filters generated with different ratios between $\sigma$ and $\lambda$. Left: $\sigma=0.1 \lambda$; Middle: $\sigma=0.3 \lambda$; Right: $\sigma=0.9 \lambda$. }
\label{fig2}
\end{figure*}

The $\sigma$ can be determined based on a reasonable relationship with $\lambda$. We show the Gabor filters with different ratios between the $\lambda$ and $\sigma$ to elucidate empirically such a relation. As shown in Figure \ref{fig2}, Gabor filters generated using $\sigma=0.1 \lambda$, $\sigma=0.3 \lambda$, and $\sigma=0.9 \lambda$, aim to extract wave components that are of wavelength shorter, equal and larger than one wavelength, respectively. 
To extract wave characteristics that can be fully represented by one wavelength of desired signals and avoid artifacts caused by the adjacent plane waves, we fix the relationship between $\lambda$ and $\sigma$ for each data scenario in this paper. 

As expected, by using the prior knowledge of the desired signals to be extracted to initialize and update the network, the network is expected to extract big portions of the desired signals at the start of the training and obtain the optimal denoising results within only several epochs.

\begin{figure*}[!t]
\centering
\includegraphics[width=5.5in,height=3.0in]{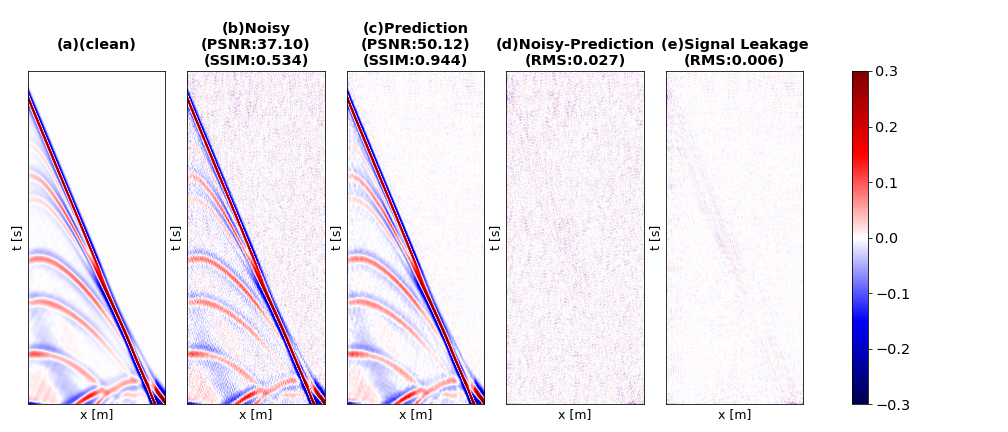}
% \captionsetup[figure]{font=Large}
\caption{The trained model applied to a synthetic dataset with field noise added.  (a)-(e) are the clean, noisy, prediction, the difference between the noisy and the prediction, and the difference between the clean and the prediction, respectively. }
\label{fig:syn-bas}
\end{figure*}

\begin{figure*}[!t]
\centering
\includegraphics[width=5.3in,height=4.8in]{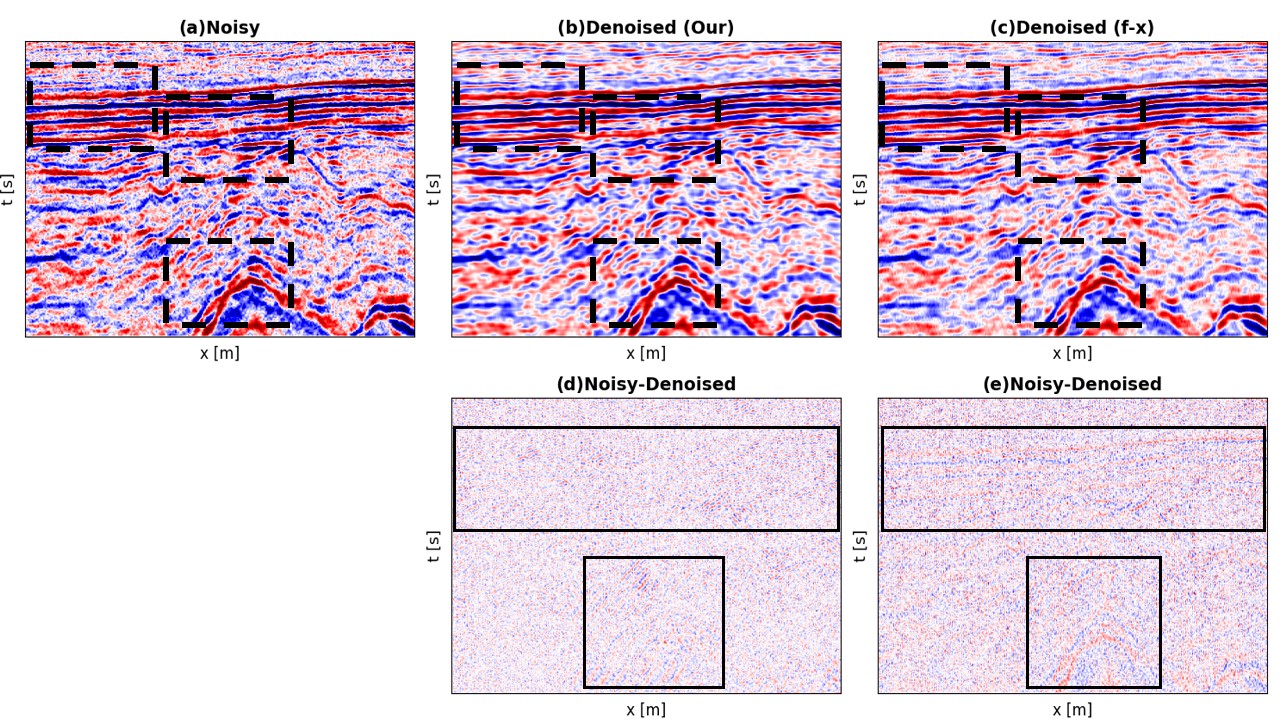}
% \captionsetup[figure]{font=Large}
\caption{The trained model applied to the field dataset with pseudo-random noise. Top: denoised results from the proposed scheme and the f-x procedure; Bottom: the corresponding difference between the original and denoised results.}
    \label{fig:field-bas}
\end{figure*}

\begin{figure*}[!t]
\centering
\includegraphics[width=3.9in,height=2.1in]{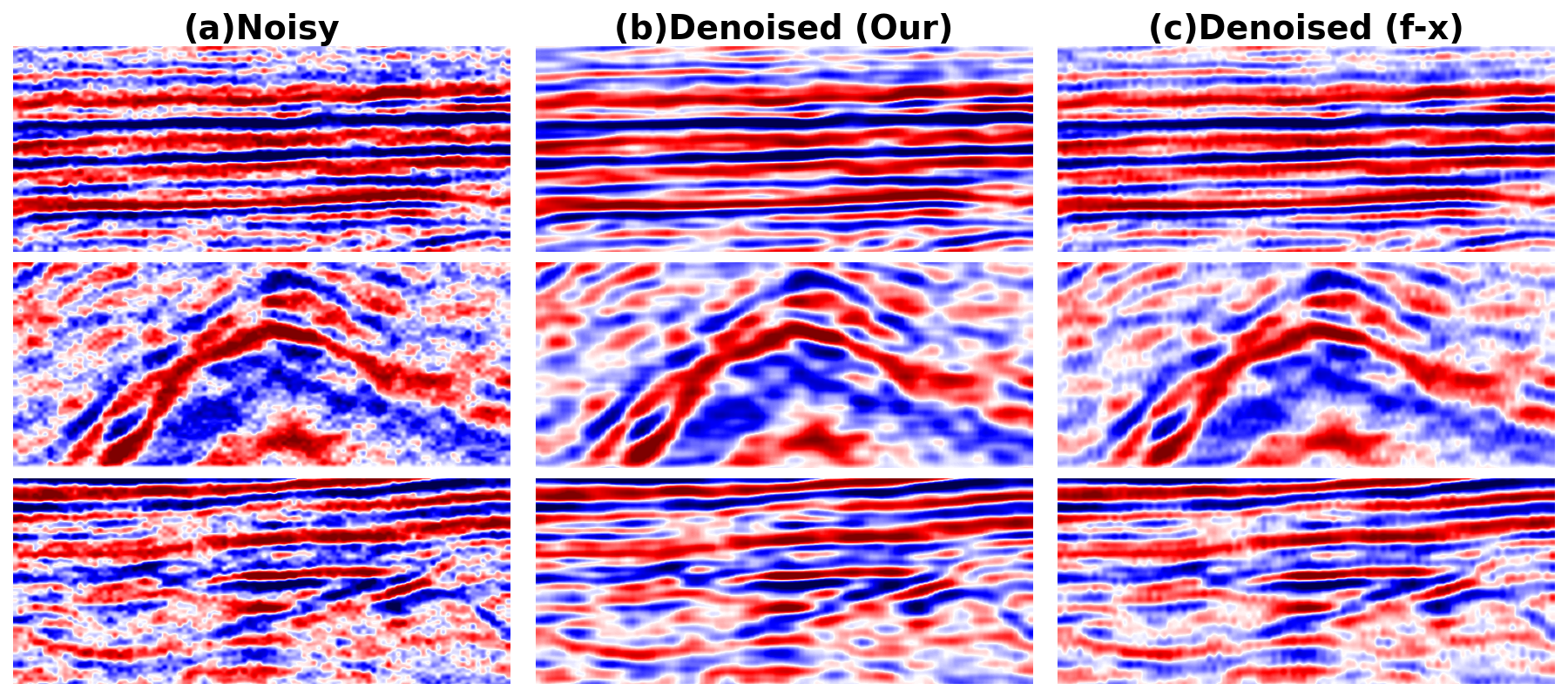}
% \captionsetup[figure]{font=Large}
\caption{A close-up of the original data (a), the proposed scheme result (b), and f-x denoised result (c), corresponding to the dashed, boxed areas shown in Figure \ref{fig:field-bas}.}
    \label{fig:zoofield-bas}
\end{figure*}

\begin{figure*}[!t]
\centering
\includegraphics[width=4.0in,height=1.7in]{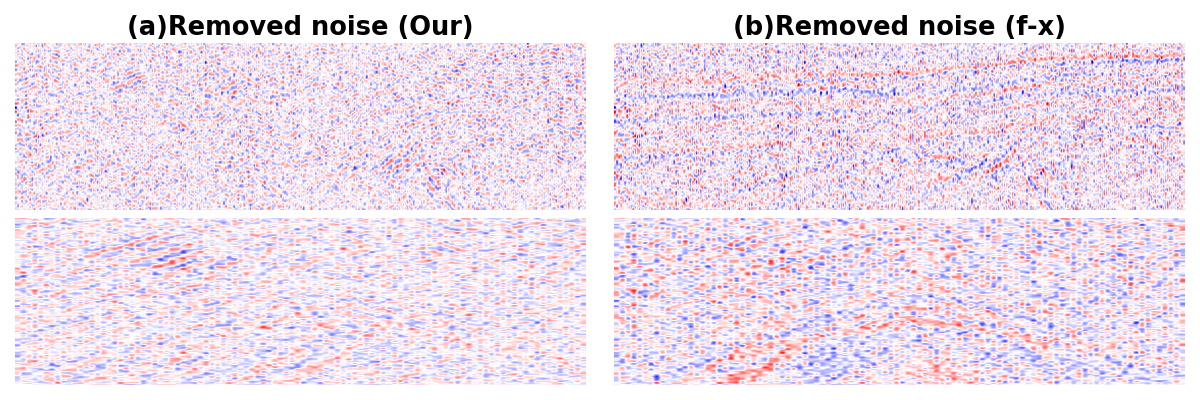}
% \captionsetup[figure]{font=Large}
\caption{A close-up of the difference between the original and denoised results of the proposed scheme (a) and f-x method (b), corresponding to the solid, boxed areas shown in Figure \ref{fig:field-bas}.}
    \label{fig:field-bas-dif}
\end{figure*}

\begin{figure*}[!t]
\centering
\includegraphics[width=5.5in,height=2.5in]{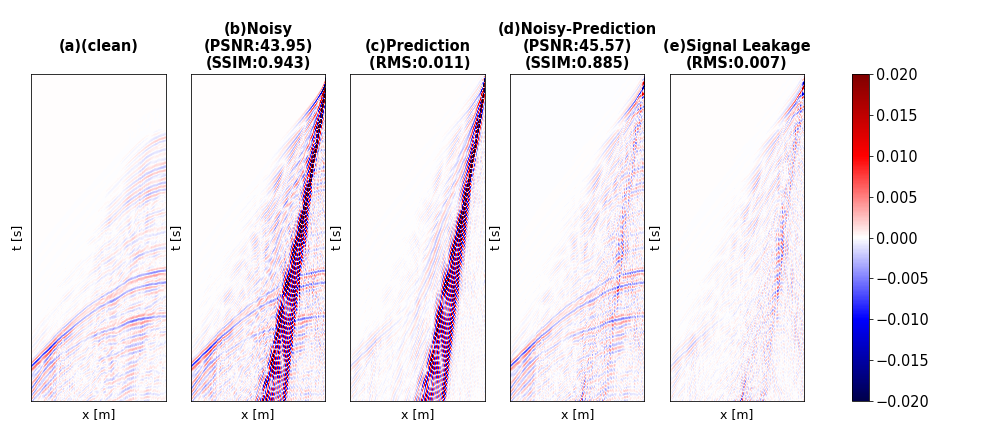}
% \captionsetup[figure]{font=Large}
\caption{The trained model applied to a synthetic dataset with ground roll.  (a)-(e) are the clean, noisy, prediction, the difference between the noisy and the prediction, and the difference between the clean and the denoised result, respectively.}
    \label{fig:syn-gr}
\end{figure*}

\begin{figure*}[!t]
\centering
\includegraphics[width=6.0in,height=4in]{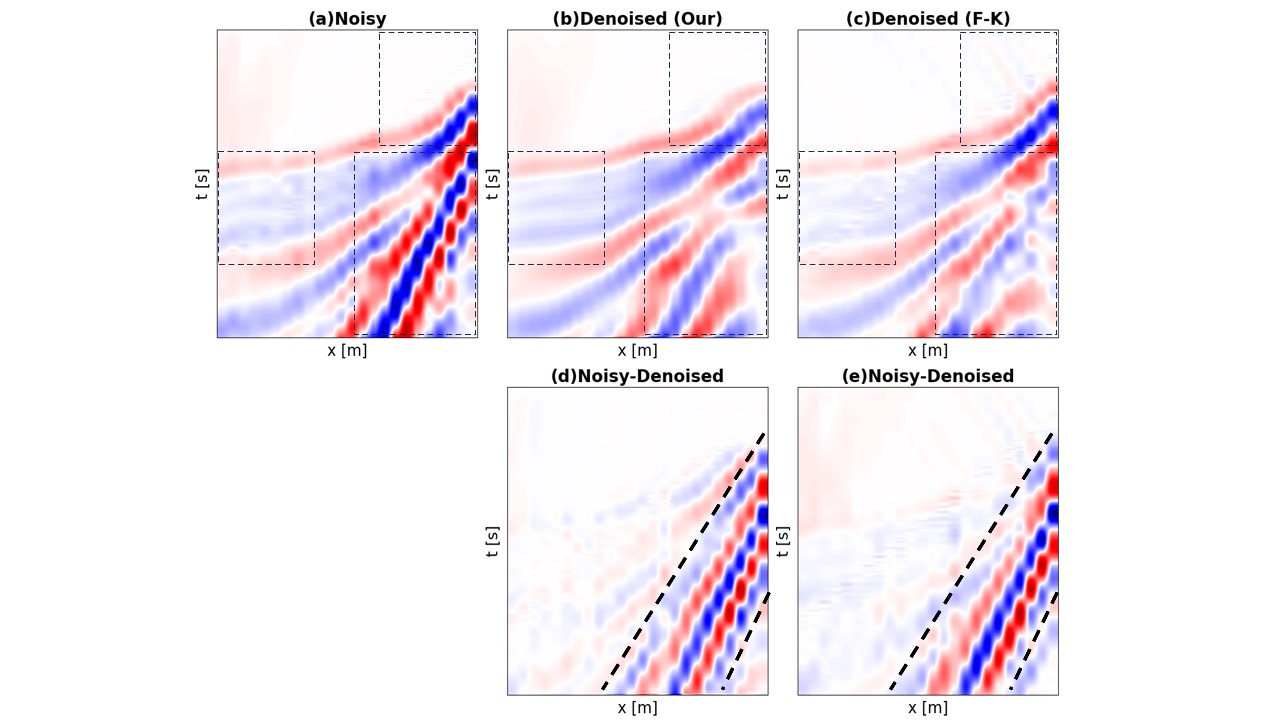}
% \captionsetup[figure]{font=Large}
\caption{The trained model applied to the field dataset with ground roll. Top: denoised results from the proposed scheme and the F-K filter; Bottom: the corresponding difference between the original and denoised results.}
    \label{fig:field-gr}
\end{figure*}

\begin{figure*}[!t]
\centering
\includegraphics[width=4in,height=3.2in]{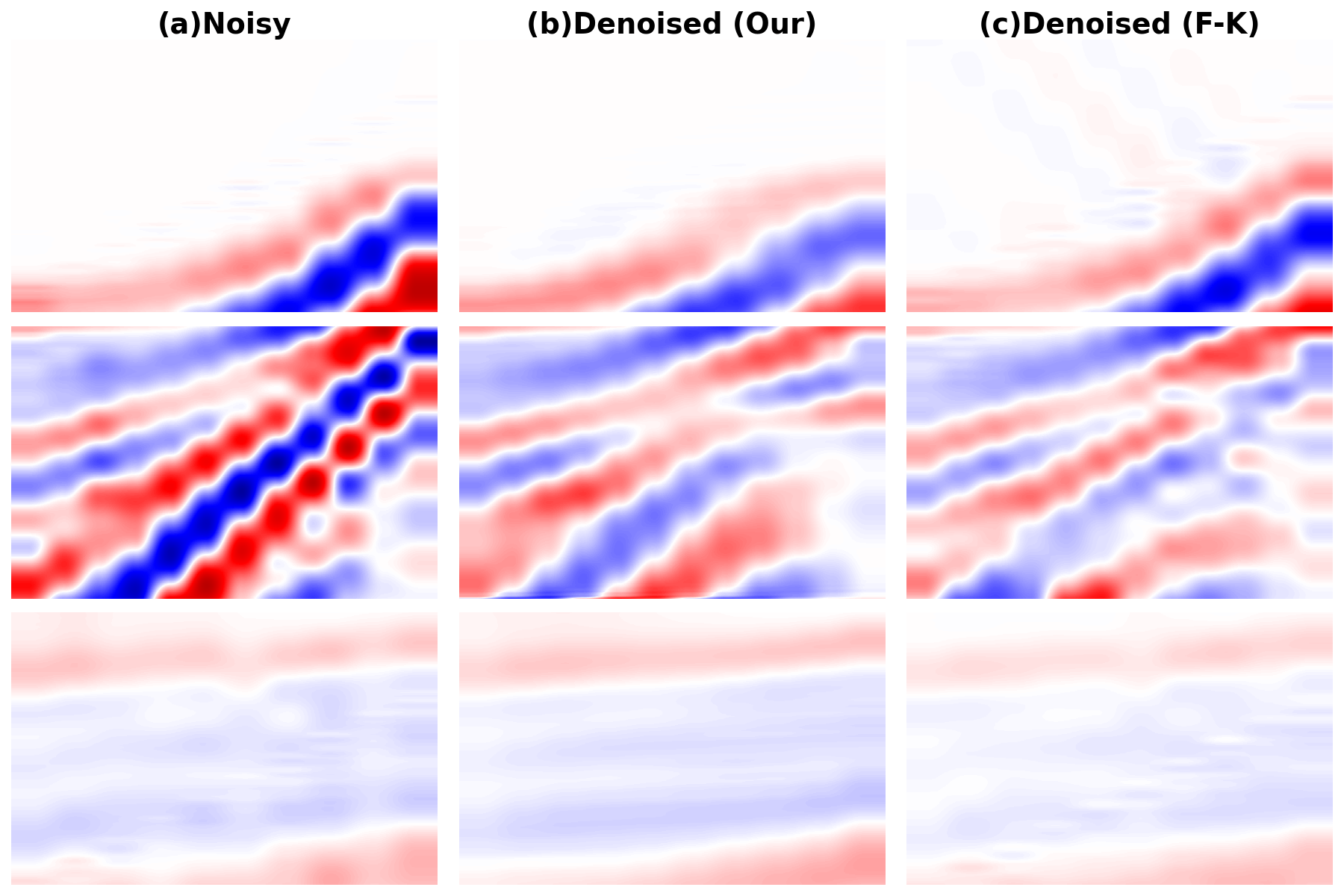}
% \captionsetup[figure]{font=Large}
\caption{A close-up of the original data (a), the proposed scheme result (b), and F-K denoised result (c), corresponding to the boxed areas shown in Figure \ref{fig:field-gr}.}
    \label{fig:zoofield-gr}
\end{figure*}

\section{Numerical Examples}
We apply the procedure to suppress two often-seen noise types in seismic data: pseudo-random noise and ground roll. In the pseudo-random noise case, we aim to extract useful signals directly from noisy data. In ground roll case, we needed to treat our synthetic and field data examples differently, as the data features are different. To be specific, in the ground roll synthetic example, we focused on maintaining the ground roll as its energy was much stronger than the signal; therefore, the residual between the prediction of the network and the noisy data are the denoised data. In the ground roll field case, the seismic shot gathers are dominated by reflections; therefore, the extracted signals are the reflections directly.

The difference between pseudo-random noise and ground roll is that ground roll is direction-oriented. Therefore, except for the wavelength required to be initialized and constrained during training in pseudo-random noise suppression case, the angle also needs to be initialized and constrained during training in ground roll removal case.

\subsection{Pseudo-random noise removal}
In this section, the proposed procedure is tested to separate effective signals from seismic data containing pseudo-random noise and thus obtain denoised results. We need to emphasize that, here, pseudo-random noise is extracted from field data \cite{liu2013noncausal,birnie2021potential} and then injected into synthetic data, as our intention is to consider removing more realistic noise. Since field data do not contain purely random noise, we refer to it as pseudo-random noise.

\textbf{(1) Synthetic data}

\textbf{Data creation}: The SEG Hess Vertical Transversely Isotropic (VTI)  model is used to acoustically model clean synthetic data. The added noise is extracted from the field data contaminated mainly by pseudo-random noise \cite{liu2013noncausal,birnie2021potential}. The summation of clean data and the noise provides us with noisy data that can be both the target and the input to our self-supervised training of the network. We compare the denoised data with the clean ones.

\textbf{Training parameters}: To use prior information to control the Gabor filters for denoising, the basic characteristics of the signals should be analyzed first. The signal components in the data have a wavelength between the range [23,37] grid points; therefore, the $\lambda$ is initialized 
by a random value in this range and constrained within this range
during training. We set $\sigma=0.23\lambda$. For this case,
there is no preference on the direction of desired signals, e.g. reflections, so we allow the range in the angle $\theta$ to extend between [$-\pi$,$\pi$]. $\alpha$ and $\psi$ are randomly initialized from a normal distribution with a mean of 0 and variance of 1. $\gamma$ is randomly initialized from the range [1.4,2.9]. The network is only trained for 3 epochs. The parameters of the Gabor and the network are shown in Table \ref{tab:para}.

\textbf{Results}: Figure \ref{fig:syn-bas} shows the denoising results of the network for the synthetic dataset with noise extracted from field data, alongside the observed Peak Signal-to-Noise Ratio (PSNR) and Structural Similarity Index Measure (SSIM). In the noisy data (Figure \ref{fig:syn-bas}(b)), the reflections are masked by the field data noise, resulting in a PSNR of 37.1 dB and an SSIM of 0.534.
The prediction of the network (Figure \ref{fig:syn-bas}(c)) looks cleaner, resulting in a higher PSNR of 50.12 dB and a higher SSIM of 0.944. The difference between the noisy and the prediction (Figure \ref{fig:syn-bas}(d)) and the difference between the clean and the prediction (Figure \ref{fig:syn-bas} (e)) indicate that the field data noise has been reduced with a subtle amount of signal leakage mainly for the early arrivals. 
The extraction of reflections is reasonably achieved, while the reduction of the noise has focused on the more higher-frequency components of the pseudo-random noise, governed by the range in the wavelength we imposed.

\textbf{(2) Field data}

\textbf{Data}: The field data (Figure \ref{fig:field-bas} (a)) are given by a 2D section of post-stack seismic data contaminated by random looking noise, acquired from a land acquisition in China and has been studied previously in \cite{birnie2021potential}. The time sampling interval is 2 ms. 

\textbf{Training parameters}: The desired signal components in the field data have a wavelength ranging between [21,35] grid points; therefore, the $\lambda$ is initialized 
by a random value in this range and constrained within this range
during training. We set $\sigma$ to be $\sigma=0.3\lambda$. Again, we do not constrain angles considering no preference in the direction in the signals, therefore, $\theta$ is randomly initialized from the range [$-\pi$,$\pi$]. $\alpha$ and $\psi$ are randomly initialized from a normal distribution with a mean of 0 and variance of 1. $\gamma$ is randomly initialized from the range [0.5,2]. The network is only trained for 4 epochs. The parameters of the Gabor and the network are shown in Table \ref{tab:para}.

\textbf{Results}: Figure \ref{fig:field-bas} illustrates the denoising results of the proposed method, and compares them with those obtained with a conventional random noise removal method often used in industry, f-x deconvolution. The raw data are shown in Figure \ref{fig:field-bas} (a). Stronger coherency of subsurface structures exists in both denoised results (Figure \ref{fig:field-bas} (b) and (c)), compared with the noisy data (Figure \ref{fig:field-bas} (a)). The differences (Figure \ref{fig:field-bas} (d) and (e)) between the noisy data and the two denoised results show that both methods can remove big portions of the noise, but
our method suffers less signal leakage than the f-x method, as highlighted by the black solid boxes. A zoomed-in comparison of three areas of interest highlighted by the black dashed boxes in Figure \ref{fig:field-bas} (a), (b), and (c), and two areas of interest highlighted by the black solid boxes in Figure \ref{fig:field-bas} (d) and (e), are shown in Figure \ref{fig:zoofield-bas} and Figure \ref{fig:field-bas-dif}, respectively, demonstrating the superiority of our method in removing pseudo-random noise, compared to the f-x method. 
\subsection{Ground roll removal}
Next, we evaluate the effectiveness of the proposed method in isolating effective reflections and ground roll from both synthetic and field data. In the synthetic data case, we focus on extracting ground roll components from the noisy data, thereby obtaining denoised results that are derived from subtracting the predicted ground roll from the noisy data.  Conversely, in the field data case, the objective is to extract reflections directly from noisy data. 

\textbf{(1) Synthetic data}

\textbf{Data creation}: A 2D slice of the SEAM Arid model is used to acoustically model the clean synthetic data while the same model with the area below 600 m set to a homogeneous half-space is used to elastically model the ground roll. Adding the ground roll to the clean data results in the noisy data, which again becomes the input and the target of our network. The data have a time sampling interval of 4 ms and a receiver spacing of 10 m.

\textbf{Training parameters}: To isolate the ground roll using the prior setting for the Gabor filter, we first analyze the basic characteristics of the ground roll. The ground roll in the noisy data has a wavelength of range [6,14] grid points; therefore, the $\lambda$ is initialized 
by a random value in this range and constrained within this range
during training. We set $\sigma$ to be $\sigma=0.3\lambda$. The slope of the ground roll is generally identifiable from the range [$0.1\pi,0.11\pi$], and thus, we initialize and constrain $\theta$ in the training stage within this slope range, according to the largest and smallest slope of the ground roll. $\alpha$, $\psi$, and $\gamma$ are randomly initialized from a normal distribution with a mean of 0 and a variance of $1$, $1$, and $1\text{e-}2$, respectively. The network is only trained for 2 epochs. The parameters of the Gabor and the network are shown in Table \ref{tab:para}.

\textbf{Results}: Figure \ref{fig:syn-gr} shows the denoising results of the network for the synthetic dataset with ground roll. The clean data are shown in Figure \ref{fig:syn-gr}(a). Some of the reflections in the noisy data (Figure \ref{fig:syn-gr}(b)) are badly contaminated by ground roll. Thanks to the Gabor angle limitations, most of the ground roll is predicted by the network in Figure \ref{fig:syn-gr}(c). Compared to the noisy data with a PSNR of 43.95 dB, the denoised result (difference between the noisy data and the prediction) shown in Figure \ref{fig:syn-gr}(d) contains much less noise and has a higher PSNR of 45.57 dB. The difference between the clean and the denoised result shown in Figure \ref{fig:syn-gr} (e) indicates some amount of signal leakage, but generally little leakage of reflection energy, and such leakage are focused on the parts of the reflection energy with large moveout (large offset). This large-offset signal leakage is related to the close slope gap between the ground roll to be removed and the large-offset signals. It is important to note that the ground roll is highly aliased, but the network managed to predict it well as a result of constraining the slope information.

\textbf{(2) Field data}

\textbf{Data}: The field data are collected at the Mirandola site in Italy and previously studied for S-wave velocity reconstruction \cite{zhang2021rayleigh}. The dataset includes a total 39 shot gathers of 400 × 24, with a time sampling interval  of 0.25 ms and a receiver spacing of 1 m. To avoid the bias of the network to either reflections or ground roll, a trace-wise normalization processing is applied to the raw field data, in order to ensure the balanced energy between reflections and ground roll.

\textbf{Training parameters}: The data are dominated by reflection energy especially in the far offsets, as compared to the ground roll. Therefore, the Gabor kernels are designed to extract the reflections. The reflections in the noisy data have a wavelength of range [4,124] grid points; therefore, the $\lambda$ is initialized 
by a random value in this range and constrained in this range
during training. We set $\sigma$ to be $\sigma=0.07\lambda$.  The slope of the reflections is generally identifiable from the range [$0.51\pi,0.64\pi$], and thus, we initialize $\theta$ before training and also constrain $\theta$ in training within this slope range, according to the largest and smallest slopes of the reflections. $\alpha$ and $\psi$ are randomly initialized from a normal distribution with a mean of 0 and a variance of $1$. $\gamma$ is randomly initialized from the range [4,14]. The network is only trained for 2 epochs. The parameters of the Gabor and the network are shown in Table \ref{tab:para}.

\textbf{Results}: Figure \ref{fig:field-gr} illustrates the denoising results of the proposed method for ground roll removal on field data, and we compare them with those obtained with a conventional ground roll removal method, known as F-K filter. In the raw data (Figure \ref{fig:field-gr} (a)), we can see that the ground roll severely contaminates the reflections, especially at near offsets. Observed from the denoised results (Figure \ref{fig:field-gr} (b) and (c)) and the difference between noisy and denoised data 
(Figure \ref{fig:field-gr} (d) and (e)), both methods are able to 
suppress most of the ground roll, with slight and comparable signal leakage. A zoomed-in comparison of the two denoising results, as highlighted by the black dashed boxes in Figure \ref{fig:field-gr} (a), (b), and (c), can be found in Figure \ref{fig:zoofield-gr}. 
We can observe that the proposed method results in more continuous reflections and fewer artifacts than the F-K filter, indicating the superiority of our method in removing ground roll, compared to the F-K filter method. 

\begin{table*}[t]\centering
% \captionsetup{font=small}
\caption{Training parameters}
  \begin{tabular}{|p{2.0cm}|p{2.3cm}|p{2.3cm}|p{2.3cm}|p{2.3cm}| }
\hline
   & PR synthetic &PR field &GR synthetic &GR field \\
\hline
Patch size & 512x256 &512x256 & 640x512 & 400x24\\
Training & 320   & 320 & 60 & 39\\
Layers  & 20 & 20& 20 & 20\\
Batch size & 4 & 4 & 4  & 4\\
Initial filters  & 32 & 32  & 32 & 32\\
Kernel size & 23 & 23 & 23 & 11\\
LR   & 1e-3 & 1e-3 & 1e-3 & 1e-3\\
Loss   & MSE & MSE & MSE & MSE\\
Epochs   & 3 & 4 & 2 & 2\\
$\alpha$ & $\mathcal{N}(0,1)$ &  $\mathcal{N}(0,1)$ & $\mathcal{N}(0,1)$  & $\mathcal{N}(0,1)$ \\
$\theta$ & $\mathcal{U}(-\pi,\pi)$ & 
$\mathcal{U}(-\pi,\pi)$& $\mathcal{U}(0.1\pi,0.11\pi)$ & $\mathcal{U}(0.51\pi,0.64\pi)$ \\
$\lambda$  & $\mathcal{U}(23,37)$ & $\mathcal{U}(21,35)$  & $\mathcal{U}(6,14)$ & $\mathcal{U}(4,124)$\\
$\sigma$& $0.23\lambda$ & $0.3\lambda$ & $0.3\lambda$ & $0.07\lambda$\\
$\psi$   & $\mathcal{N}(0,1)$ & $\mathcal{N}(0,1)$ & $\mathcal{N}(0,1)$ & $\mathcal{N}(0,1)$\\
$\gamma$  & $\mathcal{U}(1.4,2.9)$ & $\mathcal{U}(0.5,2)$ & $\mathcal{N}(0,1\text{e-}2)$ & $\mathcal{U}(4,14)$\\
 \hline
Note:& \multicolumn{2}{c|}{$\mathcal{U}:$  Uniform distribution} 
        & \multicolumn{2}{c|}{$\mathcal{N}:$  Normal distribution} \\
 \hline
     & \multicolumn{2}{c|}{$PR:$  Pseudo-random} 
        & \multicolumn{2}{c|}{$GR:$  Ground roll} \\
 \hline
 %     & \multicolumn{2}{c|}{$PR:$  Pseudo-random} 
 %        & \multicolumn{2}{c|}{$GR:$  Ground roll} \\
 % \hline
\end{tabular}
\label{tab:para}
  % \caption{Foo bar}
\end{table*}
% \section{Discussion}

\section{Conclusions}
We proposed a Gabor-based learnable sparse representation network to suppress seismic noise in a self-supervised manner. This is achieved by training an unrolled network initialized and updated by learnable Gabor filters, with the same noisy data as both inputs and targets and with much fewer learnable parameters and training epochs. By incorporating the prior knowledge of the desired signals to be extracted in the design of Gabor filters, the network allowed for the removal of various noise types by only modifying the initialization and constraining the updates of the six parameters related to the Gabor filter. The application of the method on seismic data contaminated with two different noise types has illustrated the effectiveness of the procedure in suppressing seismic noise in a self-supervised manner. The ability to control the filter will potentially allow us to remove more seismic noise types, such as speckle, swell, and trace-wise noise.

\section{Acknowledgment}
The authors thank King Abdullah University of Science  Technology (KAUST) Seismic Wave Analysis Group and the DeepWave sponsors for supporting this research and also, in particular, Dr. Claire Birnie, Dr. Yuanyuan Li, and Dr. Zhendong Zhang for insightful discussions. For computer time, this research used the resources of the Supercomputing Laboratory at KAUST in Thuwal, Saudi Arabia. 
\newpage
\bibliographystyle{plain}  

\bibliography{n2vPaper}
\end{document}